\documentclass[showpacs, amsmath, amssymb, amsthm, aps, jmp, thmsb]{revtex4}

\usepackage{color}
\usepackage{float}
\usepackage{amsmath}
\usepackage{amssymb}
\usepackage{amsfonts}
\usepackage{latexsym}
\usepackage{mathrsfs}
\usepackage[dvipdfm]{graphicx}

\newtheorem{proposition}{Proposition}
\newtheorem{theorem}{Theorem}
\newtheorem{lemma}{Lemma}
\newtheorem{corollary}{Corollary}

\newenvironment{proof}[1][Proof]{\textbf{#1.} }{\ \rule{0.5em}{0.5em}}
\newcommand{\be}{\begin{eqnarray}}
\newcommand{\ee}{\end{eqnarray}}
\def\({\left(}
\def\){\right)}

\newcommand{\braket}[1]{\left\langle #1 \right\rangle}

\newcommand{\bra}[1]{\langle #1 |}
\newcommand{\ket}[1]{| #1 \rangle}

\newcommand{\Tr}{\mathrm{Tr}}
\newcommand{\sla}[1]{\rlap{\kern .15em /}#1}
\newcommand{\s}{\hspace{5mm}}

\renewcommand{\theequation}{\arabic{section}.\arabic{equation}}

\begin{document}
\title{Exchange Symmetry and Multipartite Entanglement}
\author{
Tsubasa Ichikawa,
Toshihiko Sasaki,
Izumi Tsutsui
and
Nobuhiro Yonezawa
}

\affiliation{
High Energy Accelerator Research Organization (KEK),
Tsukuba, Ibaraki 305-0801, Japan
}

\date{Oct 10, 2008}

\begin{abstract}
Entanglement of multipartite systems is studied based on exchange symmetry under the permutation group $S_N$.  
With the observation that symmetric property under the exchange of two constituent states and their separability are intimately linked, we show that anti-symmetric (fermionic) states are necessarily globally entangled, while symmetric (bosonic) states are either globally entangled or fully separable and possess essentially identical states in all the constituent systems.  It is also shown that there cannot exist a fully separable state which is  orthogonal to all symmetric states, and that full separability of states does not survive under total symmetrization unless the states are originally symmetric.
Besides, anyonic states permitted under the braid group $B_N$ should also be globally entangled.
Our results reveal that exchange symmetry is actually sufficient for pure states to become globally entangled or fully separable.
\end{abstract}

\pacs{03.65.-w, 03.65.Ud, 03.67.-a}
\keywords{quantum mechanics, entanglement and quantum nonlocality, quantum information}

\maketitle

\section{Introduction}
One of the prominent properties of quantum states is entanglement \cite{EPR35}, which is an essential element underlying the 
impossibility of emulating genuinely quantum phenomena based on classical local realism \cite{Bell64}. 
This impossibility has been experimentally confirmed in various physical systems \cite{ADR82,WJSWZ98,RKMSIMW01,Sakai06} over the last few decades, and now we are taking up the issue of non-local realism \cite{Leggett03,GPKBZAZ07}. 
Besides its significance in the foundation of quantum mechanics, entanglement provides an indispensable resource in 
many branches of quantum information science \cite{Peres93, NC00}. 
In spite of the vital importance in these areas, entanglement defies our understanding both conceptually and theoretically, especially in multipartite systems.   This is due to the inexplicably rich and intricate structure of multipartite quantum states, which results in the exponential increase in the number of ingredients of entanglement  when the number $N$ of the constituent systems becomes large.  For example, it is known that for $N = 2$, qubit (level $n = 2$) systems
have only a single class of entanglement ordered by Schmidt coefficients \cite{Peres93, NC00}, while for $N = 3$ the systems can have many classes of entanglement \cite{Miyake03}.   The obstacle for grasping entanglement 
is also seen in the construction of entanglement measures \cite{RPMK07} for multipartite systems, where, for instance, the scalable measure proposed by Meyer and Wallach \cite{MW02} (see also \cite{B03})  has been shown, despite its feasibility,  to be flawed for quantifying the entanglement globally \cite{LMBSAGIIZ07}.

In physics, symmetry is a fundamental tool to analyze systems as well as to construct models without dealing with the insignificant detail of the structure \cite{Weyl31}.  The practical use of symmetry has been seen, {\it e.g.}, in obtaining the spectrum of the hydrogen atom
by the hidden $O(4)$ symmetry \cite{Pauli26,Schiff55}, or in classifying relativistic particle states based on the Lorentz group. 
One may then expect that a similar utility of symmetry will be found in the analysis of multipartite entangled systems as well.  
Among the various symmetries available to multipartite systems, the most fundamental is perhaps exchange symmetry, which in particle systems characterizes states of bosons, fermions, parafermions \cite{Green53,KT62,DHR70} and anyons \cite{LM77,Wilczek82,Kitaev97}.   In fact, exchange symmetry has been shown to be a key for realizing independence (or more precisely the i.i.d.~property) of subsystems in a large system \cite{KR05,Renner07},
which is important  for the validity of local experiments to infer the global property of the system.

The aim of the present paper is to demonstrate that exchange symmetry is indeed quite useful to pin down the entanglement property of multipartite states.   After observing an intimate link between the symmetric property under the exchange of two constituent states and their separability, 
we shall see that fermionic (anti-symmetric) states are always globally entangled for general $N$ and $n$, while bosonic (symmetric) states are either globally entangled or fully separable (Theorem 1).   Anyonic states share the same entanglement property with the fermionic states.
We also show that symmetric states are severely restricted with respect to the full separability of the subsystems. 
Some of the previous studies relevant to the subject, though not exploiting exchange symmetry as we do here, may be found in \cite{SCKLL01,ESBL02,GMW02,Bravyi03,WEGM04,SGDM03,GM04a,GM04b,GM05,NV07,DB07,HMMOV08}.

This paper is organized as follows.  In Sec.~II, we provide necessary settings for our arguments of multipartite entanglement, and 
thereby present our first lemma on state decompositions.  Permutation is introduced in Sec.~III to obtain the second lemma, which gives a certain restriction to states obeying permutation symmetry.  Combining the two lemmas, in Sec.~IV we present Theorem 1.   The symmetric states are further analyzed with regard to the full separability in Sec.~V, where our results are given by two propositions. Sec.~VI is dedicated to our conclusion and discussions.

\section{Decomposition of multipartite States}
\setcounter{equation}{0}

Consider the Hilbert space $\mathscr{H}$ given by the tensor product  of the constituent Hilbert spaces $\mathscr{H}_i = \mathbb{C}^n$, $i = 1, \ldots, N$,  for some positive integer $n$. 
Each space $\mathscr{H}_i$ is equipped with a set of orthonormal basis states,  
\begin{equation}
\label{basis}
\ket{0}_i,\; \ket{1}_i,\; \ldots,\; \ket{n-1}_i, \qquad i = 1, 2, \ldots, N,
 \end{equation} 
with which a state $\ket{\psi}_i \in \mathscr{H}_i$ can be expanded as 
\begin{equation}
\label{expansion}
\ket{\psi}_i = \sum_{a=0}^{n-1} c_{a}\ket{a}_i,
\qquad c_{a} \in \mathbb{C}.
\end{equation} 
The total Hilbert space $\mathscr{H}$ is spanned by the set of $N$-fold direct product states, 
  \begin{equation}
   \bigotimes_{i=1}^N\ket{a_i}_i, \s a_i \in \{0, 1, \ldots, n-1\}.
  \end{equation}

To discuss subsystems of the Hilbert space $\mathscr{H}$, we consider the set of labels of the constituents,
\be
{\cal N}:=\{1,2,\cdots,N\},
\ee
and introduce a partition ${\mathscr A}$ consisting of exclusive subsets ${\mathscr A}=\{{\mathscr A}_k\}_{k=1}^s$ covering ${\cal N}$, {\it i.e.},
\be
\bigcup_{k=1}^s{\mathscr A}_k={\cal N}
\quad
{\rm and}
\quad
{\mathscr A}_j\cap {\mathscr A}_k=\emptyset
\quad
{\rm for}
\quad
j\neq k.
\label{defset}
\ee
Each subset ${\mathscr A}_k$ determines the tensor product Hilbert space ${\mathscr H}({\mathscr A}_k):=\bigotimes_{i\in{\mathscr A}_k}{\mathscr H}_i$ as a subsystem of ${\mathscr H}$. 
In order to treat the labels of the basis states in ${\mathscr H}({\mathscr A}_k)$ collectively, we also introduce
\be
\alpha_k:=\{a_i\,|\,i\in{\mathscr A}_k\},
\label{alpha}
\ee
and thereby denote the basis states in ${\mathscr H}({\mathscr A}_k)$ as
\be
\ket{\alpha_k}:=\bigotimes_{i\in{\mathscr A}_k}\ket{a_i}_i.
\label{alphaket}
\ee
A state $\ket{A_k}\in{\mathscr H}({\mathscr A}_k)$ is then written as
\be
\ket{A_k} =\sum_{\alpha_k}A_{\alpha_k}\ket{\alpha_k},
\qquad
A_{\alpha_k}\in{\mathbb C},
\label{statepartial}
\ee
where the summation of $\alpha_k$ is 
over all $a_i=0,1,\cdots,n-1$ for $i\in{\mathscr A}_k$ appearing in $\alpha_k$. For convenience, we assume that in each of the subsystems the state is normalized:
\be
\braket{A_k|A_k}=1.
\ee

By choosing a partition ${\mathscr A}=\{{\mathscr A}_k\}_{k=1}^s$ properly, and taking the direct product of the state $\ket{A_k}$ in (\ref{statepartial}) for all $k$, we can express any state $\ket{\psi}\in{\mathscr H}$ in the product form of subsystems,
\be
\ket{\psi}=\bigotimes_{k=1}^s\ket{A_k}.
\label{statedp}
\ee
Note that the expression of the product (\ref{statedp}) is not unique for a given state $\ket{\psi}\in{\mathscr H}$, since it depends on the choice of the bases in the subsystems used in the expansion.  Moreover, even under the same choice, the state may still admit different product expressions having distinct partition number $s$ due to the ambiguity in assigning the partition for the state.   Among all possible $s$ in those admitted partitions for a given $\ket{\psi}$, there exists the maximal partition number which gives the number of subsystems into which the state $\ket{\psi}$ can be decomposed at most.   Adopting the notation $M(\psi)$ for the maximal partition number of the state $\ket{\psi}$, we say that the state with $M(\psi) = d$ is a {\it $d$-fold direct product state}.  
If $M(\psi)=N$, the state $\ket{\psi}$ is called {\it fully separable}, and if $M(\psi)=1$, the state is {\it globally entangled}.  Otherwise, the state is {\it partially entangled}.   To the former two extreme cases, we introduce the space ${\mathscr T}$ of all fully separable states and the space ${\mathscr G}$ of all globally entangled states, {\it i.e.},
    \begin{equation}
     \mathscr{T}:=\big\{\, \ket{\psi} \;\big{\vert} \; M(\psi) = N \big\},
      \qquad
      \mathscr{G}:=\big\{\, \ket{\psi} \;\big{\vert} \; M(\psi) = 1 \big\}.
    \end{equation}
    
Given two partitions, ${\mathscr A}=\{{\mathscr A}_k\}_{k=1}^s$ and ${\mathscr B}=\{{\mathscr B}_l\}_{l=1}^t$, we can construct another partition ${\mathscr C}=\{{\mathscr C}_m\}_{m=1}^u$ by collecting non-empty intersections of ${\mathscr A_k}$ and ${\mathscr B}_l$ for all $k$ and $l$, {\it i.e.},
if $m = m(k, l) \in \{1, 2, \ldots, u\}$ labels such nonempty subsets formed by ${\mathscr A_k}\cap{\mathscr B}_l$,  we have
\be
{\mathscr C}=\{{\mathscr C}_m\}_{m=1}^u,
\qquad
{\mathscr C}_m={\mathscr A}_k\cap{\mathscr B}_l\neq\emptyset,
\label{meetab}
\ee
where obviously $1\le u\le st$.  We denote the partition defined this way symbolically by ${\mathscr C} = {\mathscr A}\cap {\mathscr B}$.

We can also define the complement $\bar{\mathscr A}_k$ of ${\mathscr A}_k$ as a set satisfying
\be
{\mathscr A}_k\cup\bar{\mathscr A}_k={\cal N},
\qquad
{\mathscr A}_k\cap\bar{\mathscr A}_k=\emptyset.
\ee
Clearly, $\{{\mathscr A}_k, \bar{\mathscr A}_k\}$ qualifies as a partition consisting of the two subsets, and 
accordingly we may also employ
\be
\bar{\alpha}_k:=\{a_i\,|\,i\in\bar{\mathscr A}_k\},
\ee
for the basis states $\ket{\bar{\alpha}_k}$ in the subsystem ${\mathscr H}(\bar{\mathscr A}_k)$,
\be
\ket{\bar{\alpha}_k}:=\bigotimes_{i\in\bar{\mathscr A}_k}\ket{a_i}_i.
\ee
The partition just introduced is useful to express the density matrix associated with the subsystem ${\mathscr H}({\mathscr A}_k)$.  Indeed, from the density matrix $\rho=\ket{\psi}\bra{\psi}$ of the state $\ket{\psi}\in{\mathscr H}$, we obtain the density matrix $\rho_{\alpha_k}$ for the subsystem  ${\mathscr H}({\mathscr A}_k)$ by the partial trace over the complemental subsystem ${\mathscr H}(\bar{\mathscr A}_k)$,
\be
\rho_{\alpha_k}=\Tr_{\bar{\alpha}_k}\rho:=\sum_{\bar{\alpha}_k}\bra{\bar{\alpha}_k}\rho\ket{\bar{\alpha}_k}.
\label{rdm}
\ee
In particular, for the state $\ket{\psi}$ in the decomposed form (\ref{statedp}), we have 
\be
\rho=\bigotimes_{k=1}^s\rho_{\alpha_k},
\qquad
\rho_{\alpha_k}=\ket{A_k}\bra{A_k}.
\label{dmsep}
\ee
Now we consider the density matrix $\rho_{\beta_l} = \Tr_{\bar{\beta}_l}\rho$ of the subsystem ${\mathscr H}({\mathscr B}_l)$ defined by
another partition ${\mathscr B}=\{{\mathscr B}_l\}_{l=1}^t$.   Analogous to the subsystems ${\mathscr H}({\mathscr A}_k)$ where we use $\alpha_k$ to denote collectively the basis states, we shall use $\beta_l$ to denote the basis states in the subsystems ${\mathscr H}({\mathscr B}_l)$.  Since ${\mathscr A}=\{{\mathscr A}_k\}_{k=1}^s$ is a partition
we have $\bar{\mathscr B}_l = \cup_k ({\mathscr A}_k\cap \bar{\mathscr B}_l)$, and hence the trace of the density matrix $\rho$ over ${\mathscr H}(\bar{\mathscr B}_l)$
may also be realized by successive traces over the subsystems ${\mathscr H}({\mathscr A}_k\cap\bar{\mathscr B}_l)$ for
all $k$ as
\be
\Tr_{\bar{\beta}_l}\rho=\Tr_{\alpha_1\cap\bar{\beta}_l}\Tr_{\alpha_2\cap\bar{\beta}_l}\cdots\Tr_{\alpha_s\cap\bar{\beta}_l}\rho.
\label{pdtid}
\ee
Combining with (\ref{dmsep}), we find for $\ket{\psi}$ in (\ref{statedp}) the identity,
\be
\rho_{\beta_l}=\Tr_{\bar{\beta}_l}\rho=\Tr_{\alpha_1\cap\bar{\beta}_l}\Tr_{\alpha_2\cap\bar{\beta}_l}\cdots\Tr_{\alpha_s\cap\bar{\beta}_l}\bigotimes_{k=1}^s\rho_{\alpha_k}=\bigotimes_{k=1}^s\Tr_{\alpha_k\cap\bar{\beta}_l}\rho_{\alpha_k}.
\label{idenpt}
\ee

With this preparation, we obtain
\begin{lemma}
If a pure state $\ket{\psi}\in{\mathscr H}$ can be decomposed in two ways by partitions ${\mathscr A}=\{{\mathscr A}_k\}_{k=1}^s$ and ${\mathscr B}=\{{\mathscr B}_l\}_{l=1}^t$, 
\begin{eqnarray}
\ket{\psi}=\bigotimes_{k=1}^s\ket{A_k}=\bigotimes_{l=1}^t\ket{B_l},
\label{ways}
\end{eqnarray}
then, the state can be further decomposed into subsystems defined by the partition ${\mathscr C} = {\mathscr A}\cap {\mathscr B}$, that is, it can be written as
\be
\ket{\psi}=\bigotimes_{m=1}^u\ket{C_m},
\qquad
\ket{C_m}\in{\mathscr H}({\mathscr C}_m),
\label{ans}
\ee
where ${\mathscr H}({\mathscr C}_m)$ are the subsystems associated with the partition ${\mathscr C}=\{{\mathscr C}_m\}_{m=1}^u$ defined in (\ref{meetab}).
\end{lemma}
\begin{proof}
Consider the density matrix $\rho=\ket{\psi}\bra{\psi}$ and the reduced density matrix $\rho_{\beta_l}$ defined in (\ref{rdm}). Since the state admits the decomposition under the partition ${\mathscr A}$, we have (\ref{idenpt}).
But since $\rho_{\beta_l}=\ket{B_l}\bra{B_l}$ is a density matrix for a pure state, we have 
$\rho_{\beta_l}^2=\rho_{\beta_l}$ and, hence, from (\ref{idenpt}) and (\ref{pdtid}) used for $\beta_l$ instead of $\bar\beta_l$, we find
\be
\begin{split}
1=\Tr_{\beta_l}\rho_{\beta_l}
=\Tr_{\beta_l}\rho_{\beta_l}^2
=\prod_{k=1}^s\Tr_{\alpha_k\cap\beta_l}\(\Tr_{\alpha_k\cap\bar{\beta}_l}\rho_{\alpha_k}\)^2.
\label{dunit}
\end{split}
\ee
This shows that $\Tr_{\alpha_k\cap\beta_l}\(\Tr_{\alpha_k\cap\bar{\beta}_l}\rho_{\alpha_k}\)^2=1$ for all $k$ and $l$, because the trace of a squared density matrix has an upper bound equal to unity, and this must be attained in (\ref{dunit}).   Since the upper bound is attained if and only if the state represented by the density matrix is pure, we learn that $\Tr_{\alpha_k\cap\bar{\beta}_l}\rho_{\alpha_k}$ corresponds to a pure state. 
It follows that the state has to be a direct product of the traced part and the rest if none of the subsets, ${\mathscr A}_k\cap {\mathscr B}_l$ and ${\mathscr A}_k\cap \bar{\mathscr B}_l$, is empty.  Consequently, the state $\ket{\psi}\in{\mathscr H}$ can be decomposed into three subsystems as 
\be
\ket{\psi} = \ket{C}\otimes\ket{C'}\otimes\ket{\bar A},
\qquad
 \ket{C}\in{\mathscr H}({\mathscr A}_k\cap {\mathscr B}_l),
 \quad 
\ket{C'}\in {\mathscr H}({\mathscr A}_k\cap \bar{\mathscr B}_l),
 \quad 
\ket{\bar A}\in {\mathscr H}(\bar{\mathscr A}_k).
\label{decone}
\ee
Since this is true for all $k, l$ which generate non-empty intersections, and since the corresponding subsets 
${\mathscr A}_k\cap {\mathscr B}_l$ are exclusive from each other and comprise the partition  ${\mathscr C} = {\mathscr A}\cap {\mathscr B}$ in (\ref{meetab}), we arrive at (\ref{ans}).
\end{proof}

\section{Decomposed States with Permutation Symmetry}
\setcounter{equation}{0}
Next we discuss a consequence of permutation symmetry on the specification of states admitting a certain type of decompositions.
To this end, given an  elemant $\sigma$ of the symmetric group $S_N$ consisting of all permutations $i \to \sigma(i)$
of the elements of ${\cal N}$, we first define the corresponding linear operator $\pi_\sigma$ by the action of $\sigma$ on the label of the basis states,
   \begin{equation}
   \label{permac}
\pi_\sigma   \bigotimes_{i=1}^N\ket{a_i}_i =
      \bigotimes_{i=1}^N\ket{a_{\sigma(i)}}_i.
   \end{equation}       
Let ${\mathscr I}_\sigma$ be a set of labels of the constituents which are invariant under the action of the given $\sigma\in S_N$,
\be
{\mathscr I}_\sigma:=\{i\,|\,\sigma(i)=i\}.
\ee
The complement of ${\mathscr I}_\sigma$ in ${\cal N}$ is
\be
\bar{\mathscr I}_\sigma=\{i\,|\,\sigma(i)\neq i\}.
\ee
Observe that both ${\mathscr I}_\sigma$ and $\bar{\mathscr I}_\sigma$ are invariant under the multiple actions of $\sigma$,
\be
\sigma^p{\mathscr I}_\sigma={\mathscr I}_\sigma,
\qquad
\sigma^p\bar{\mathscr I}_\sigma=\bar{\mathscr I}_\sigma,
\qquad
\forall\, p\in{\mathbb Z}.
\ee
With respect to the actions under $\sigma^p$, one can introduce an equivalence relation between $i,j\in\bar{\mathscr I}_\sigma$ by
\be
i\sim j
\quad
\Leftrightarrow
\quad
i=\sigma^p(j),
\quad
\exists p\in{\mathbb Z},
\ee
which provides the equivalent classes  $\bar{\mathscr I}_\sigma/\sim$.
We note that $\{{\mathscr I}_\sigma,\bar{\mathscr I}_\sigma\}$ also give a partition with the corresponding subsystems ${\mathscr H}(\mathscr{I}_\sigma)$ and ${\mathscr H}(\bar{\mathscr{I}}_\sigma)$, respectively.

Given two states in different constituent spaces, $\ket{\psi}_i\in{\mathscr H}_i$ and $\ket{\varphi}_j\in{\mathscr H}_j$, 
we say that these states are {\it equipollent} and write $\ket{\psi}_i\simeq\ket{\varphi}_j$ if all coefficients of $\ket{\psi}_i$ and $\ket{\varphi}_j$  coincide up to a phase in the expansion under some fixed orthonormal bases (\ref{expansion}) in correspondence.   Namely, if we expand the states as $\ket{\psi}_i=\sum_{k=0}^{n-1}c_{k}\ket{k}_i$ and $\ket{\varphi}_j=\sum_{k=0}^{n-1}d_{k}\ket{k}_j$ with some bases $\{\ket{k}_i\}$ and $\{\ket{k}_j\}$ in the two constituent spaces which are regarded to be  correspondent to each other ({\it i.e.}, $\ket{k}_i \leftrightarrow \ket{k}_j$ for all $k$), then
\be
\ket{\psi}_i\simeq\ket{\varphi}_j
\quad
\Leftrightarrow
\quad
c_{k}=\gamma\, d_{k},
\quad
\forall\, k,
\label{equipl}
\ee
where $\gamma$ is a phase factor $|\gamma|=1$ common for all $k$.   In particular, if (\ref{equipl}) holds with
$\gamma = 1$, we say that the two states are {\it strictly equipollent}.
We now show

\begin{lemma}
Suppose that a state $\ket{\psi}\in{\mathscr H}$ satisfies
  
\noindent
  (i) $\ket{\psi}$ is an eigenstate for a given $\pi_\sigma$ with eigenvalue $\lambda$,
 \begin{equation}
\pi_\sigma\ket{\psi}=\lambda\ket{\psi},
\label{sc}
\end{equation}

\noindent
(ii)\,$\ket{\psi}$ is decomposed into states in ${\mathscr H}(\mathscr{I}_\sigma)$ and ${\mathscr H}(\bar{\mathscr I}_\sigma)$,
\be
\ket{\psi}=\ket{I}\otimes\ket{\bar{I}},
\qquad
\ket{I}\in{\mathscr H}({\mathscr I}_\sigma),
\qquad
\ket{\bar{I}}\in{\mathscr H}(\bar{\mathscr I}_\sigma),
\ee
such that $\ket{\bar{I}}$ is a direct product of states $\ket{\psi_i}_i\in{\mathscr H}_i$ in the constituent spaces,
 \be
\ket{\bar{I}}=\bigotimes_{i\in{\bar{\mathscr I}_\sigma}} \ket{\psi_i}_i.
\ee
Then the eigenvalue is unity $\lambda=1$, and the states of the constituents whose labels belong to the same equivalent class 
$\bar{\mathscr I}_\sigma/\sim$ are all equipollent, 
\be
\ket{\psi_i}_i\simeq\ket{\psi_j}_j
\quad
{\rm for}
\quad
i\sim j,
\quad
 i,j\in\bar{\mathscr I}_\sigma,
 \label{lattersecondlemma}
\ee
in the strict sense.
\end{lemma}
\begin{proof}
We first expand $\ket{\psi_i}_i$ for all $i\in\bar{\mathscr I_\sigma}$ as
\be
\ket{\psi_i}_i=\sum_{a_i=0}^{n-1}c_{a_{i}}^i\ket{a_i}_i, \qquad
c_{a_{i}}^i \in \mathbb{C},
\label{psiexpand}
\ee
and observe that the left hand side of (\ref{sc}) becomes
   \be
  \pi_\sigma\ket{\psi}=\ket{I}\otimes\bigotimes_{i\in\bar{\mathscr I}_\sigma}\(\sum_{a_i=0}^{n-1}c_{a_i}^i\ket{a_{\sigma(i)}}_i\)=\ket{I}\otimes\bigotimes_{i\in\bar{\mathscr I}_\sigma}\(\sum_{a_{\sigma(i)}=0}^{n-1}c_{a_{\sigma(i)}}^{\sigma(i)}\ket{a_{\sigma(i)}}_i\),
  \ee
because under the direct product over all elements in $\bar{\mathscr I}_\sigma$ we can exchange the coefficients within $\bar{\mathscr I}_\sigma$ freely. 
Changing the labels $a_{\sigma(i)}\rightarrow a_i$ which are dummy indices of the summations, we find that
(\ref{sc}) reads
  \be
  \pi_\sigma\ket{\psi} =\ket{I}\otimes\bigotimes_{i\in\bar{\mathscr I}_\sigma}\(\sum_{a_i=0}^{n-1}c_{a_i}^{\sigma(i)}\ket{a_i}_i\)=\lambda\ket{\psi}.
  \label{stransposed}
  \ee
Taking the inner products with $\bigotimes_{i\in\bar{\mathscr I}_\sigma}{_i\bra{a}}$ for 
$a \in \{0, 1, \ldots, n-1\}$ in (\ref{stransposed}),  we obtain
\be
\prod_{i\in\bar{\mathscr I}_\sigma}c^{\sigma(i)}_a=\lambda\prod_{i\in\bar{\mathscr I}_\sigma}c^{i}_{a}.
\label{resu}
\ee
Since the product in (\ref{resu}) is over all elements of $\bar{\mathscr I}_\sigma$, we have the identity,
\be
\prod_{i\in\bar{\mathscr I}_\sigma}c^{\sigma(i)}_{a}=\prod_{\sigma(i)\in\bar{\mathscr I}_\sigma}c^{\sigma(i)}_{a}=\prod_{i\in\bar{\mathscr I}_\sigma}c^i_{a}.
\label{trans}
\ee
 Combining (\ref{trans}) with (\ref{resu}), we find $\lambda=1$. 

Now choose an element $i \in \bar{\mathscr I}_\sigma$ and consider the state $\ket{\tilde\psi}= \ket{a}_i{}_i\!\braket{a|\psi}$ which is equivalent to $\ket{\psi}$ except that 
its constituent state $\ket{\psi_i}_i$ is replaced by $\ket{a}_i$.  
The inner product of $\ket{\tilde\psi}$ with $\ket{\psi}$ in (\ref{stransposed}) then yields
\be
c_{a}^{\sigma(i)} = c_{a}^i.
\ee
Since this is true for all $i \in \bar{\mathscr I}_\sigma$ and $a \in \{0, 1, \ldots, n-1\}$, we find (\ref{lattersecondlemma}).
\end{proof}

\section{Global Entanglement and Symmetry of States}
\setcounter{equation}{0}

In this section, we focus on multipartite states possessing particular symmetric properties under transpositions.  We first introduce the set of all transpositions ${\cal T}_N \subset S_N$ by
\be
{\cal T}_N:=\{\sigma\, |\,\sigma=(i\,j),\,\forall\, i,j\in{\cal N}\}.
\ee
Combining the Lemmas in the previous sections and applying them to transpositions, we obtain

\begin{proposition}
Suppose that there exists a state $\ket{\psi}$ decomposed as (\ref{statedp}) under a partition ${\mathscr A}$
consisting of two subsets, ${\mathscr A}=\{{\mathscr A}_1, {\mathscr A}_2\}$.
Suppose further that $\ket{\psi}$ is an eigenstate of $\pi_\sigma$,
\be
\pi_\sigma\ket{\psi}=\lambda\ket{\psi},
\label{sym}
\ee
for some  transposition $\sigma=(i\,j)\in{\cal T}_N$ such that
\be
i\in{\mathscr A}_1
\quad
{\rm and}
\quad
j\in{\mathscr A}_2.
\ee
Then the state $\ket{\psi}$ can be decomposed further by the partition ${\mathscr C}={\mathscr A}\cap{\mathscr B}$, where
${\mathscr B}=\{{\mathscr B}_1, {\mathscr B}_2\}$ is the partition defined from ${\mathscr A}$ with the elements $i$ and $j$ interchanged.
Moreover, the eigenvalue is unity $\lambda=1$ and the states $\ket{\psi_i}_i$ and $\ket{\psi_j}_j$ of $\ket{\psi}$ in the constituent spaces ${\mathscr H}_i$ and ${\mathscr H}_j$ are equipollent $\ket{\psi_i}_i\simeq\ket{\psi_j}_j$ in the strict sense.
\end{proposition}
\begin{proof}
With no loss of generality, we may consider the partition
\be
{\mathscr A}_1=\{1, \cdots, i, \cdots, M\}
\quad
{\rm and}
\quad
{\mathscr A}_2=\{M+1,\cdots, j, \cdots, N\}.
\ee
Observe that since $\pi_\sigma^2 = I$ (identity operator), we have $\lambda = \pm 1$.  Then (\ref{sym}) implies 
$\ket{\psi} = \lambda^{-1}\pi_\sigma\ket{\psi}$, that is, the state $\ket{\psi}$ admits the decomposition (\ref{statedp}) under another partition ${\mathscr B}=\{{\mathscr B}_1, {\mathscr B}_2\}$, where
\be
{\mathscr B}_1=\{1, \cdots, j, \cdots, M\}
\quad
{\rm and}
\quad
{\mathscr B}_2=\{M+1,\cdots, i, \cdots, N\}.
\ee
Lemma 1 then assures that the state $\ket{\psi}$ can be decomposed further by the partition ${\mathscr C}={\mathscr A}\cap{\mathscr B}$.
Noting that ${\mathscr A}_1\cap{\mathscr B}_2 = \{i\}$ and ${\mathscr A}_2\cap{\mathscr B}_1 = \{j\}$, we find that $\ket{\psi}$ has the 
constituent states $\ket{\psi_i}_i \in {\mathscr H}_i$ and $\ket{\psi_j}_j\in {\mathscr H}_j$ in the decomposition.   Since these constituent states belong to the  subsystem ${\mathscr H}(\bar{\mathscr I}_\sigma)$, the rest of the statements of Proposition 1 follows from 
Lemma 2.
 \end{proof}\\

This proposition implies
\begin{corollary}
If a state $\ket{\psi}$ satisfies $\pi_\sigma\ket{\psi}= -\ket{\psi}$ under a transposition $\sigma=(i\,j)$, then $\ket{\psi}$ cannot be separable with respect to the constituent subspaces ${\mathscr H}_i$ and ${\mathscr H}_j$.  On the other hand, 
if $\ket{\psi}$ satisfies $\pi_\sigma\ket{\psi}= \ket{\psi}$ under  $\sigma=(i\,j)$, then two possibilities arise: either 
$\ket{\psi}$ is not separable with respect to ${\mathscr H}_i$ and ${\mathscr H}_j$, or otherwise it 
is separable with respect to  ${\mathscr H}_i$ and ${\mathscr H}_j$ where the constituent states are strictly equipollent to each other.
\end{corollary}

To proceed, we introduce the totally symmetrized subspace,
    \begin{equation}
     \mathscr{S}:=\big\{\, \ket{\psi} \;\big{\vert} \;
      \pi_\sigma\ket{\psi}=\ket{\psi},\, \forall\,\sigma \in S_N \big\},
      \label{symdef}
    \end{equation}
and the subset,
\begin{equation}
\mathscr{D}= \mathscr{T} \cap \mathscr{S} 
\label{dspacedef}
\end{equation}
which consists of $N$-fold direct product states belonging to $\mathscr{S}$. 
Consider now a state $\ket{\psi}\in{\mathscr H}$ fulfilling
\be
\pi_\sigma\ket{\psi}=\lambda_{\sigma}\ket{\psi},
\quad
\lambda_\sigma=\pm1,
\label{bf}
\ee
for all $\sigma\in{\cal T}_N$. 
Since $\ket{\psi}$ is a simultaneous eigenstate of the linear operators of the transpositions, multiple operations of the different transpositions on the state is naturally induced. 
It is well known that a non-trivial $\ket{\psi}$ with the condition (\ref{bf}) must be in one dimensional representations of $S_N$, and from this we have 
\be
\lambda_{\sigma\tau}=\lambda_\sigma\lambda_\tau,
\qquad
\forall\,\sigma,\tau\in S_N.
\ee
There are two different one dimensional representations for $S_N$: one is $\lambda_\sigma=-1$ for all $\sigma\in{\cal T}_N$, the other is $\lambda_\sigma=1$ for all $\sigma\in{\cal T}_N$.  A state realizing the former is called {\it anti-symmetric} whereas a state realizing the latter is called {\it symmetric} \cite{Weyl31}, and in particle systems these correspond to states of fermions and bosons, respectively.
We then present

\begin{theorem}  The following two statements hold:

\noindent
(i) Any anti-symmetric state $\ket{\psi}$  is globally entangled $\ket{\psi} \in {\mathscr G}$.   

\noindent
(ii) Any symmetric state $\ket{\psi}$  is either globally entangled $\ket{\psi} \in {\mathscr G}$,  or otherwise it is
fully separable $\ket{\psi} \in {\mathscr T}$.  In the latter case, all the constituent states of $\ket{\psi}$ are strictly equipollent. 
\end{theorem}
\begin{proof}
Both of the statements follow immediately from Corollary 1 by applying it to all transpositions.  
\end{proof}\\

We note that, in terms of spaces, statement (ii) implies
\be
\mathscr{S}=\mathscr{S}\cap\(\mathscr{T}\cup\mathscr{G}\)=\mathscr{D}\cup\(\mathscr{G}\cap\mathscr{S}\).
\label{symset}
\ee
We also mention that statement (ii) allows us to characterize concisely
the space $\mathscr{D}$ of fully separable symmetric states by means of unitary operators.   To see this, we first consider unitary operators
$U(\theta^i)$ implementing $SU(n)$ transformations on the basis in $\mathscr{H}_i$, where 
$\theta^i= (\theta_1^i, \theta_2^i, \ldots,  \theta_m^i) \in \Theta$ ($m = {\dim \mathfrak{su}(n)} = n^2-1$) are the set of parameters specifying the transformations with $\Theta$ being the parameter space covering the entire $SU(n)$.  
The unitary operator implementing the $SU(n)$ transformations for all the bases in the constituent Hilbert spaces  
$\mathscr{H}_i$ can then be given by the product $\otimes_{i=1}^N U(\theta^i)$.  Since  statement (ii) assures that
all constituent states in $\mathscr{D}$ are strictly equipollent, we can use the same unitary operator for all constituent spaces
$\mathscr{H}_i$ with common  $\theta$ to generate the constituent states $\ket{\psi_i}_i$ from, say, the state $\ket{0}_i$ as $\ket{\psi_i}_i = U(\theta) \ket{0}_i$.  If this is done, then 
we have
\begin{corollary}
The space $\mathscr{D}$ of fully separable symmetric states is characterized by
\begin{equation}
\label{std}
	   \mathscr{D} =  
\left\{ \, \ket{\psi} \, \bigg\vert \,  \ket{\psi} = \bigotimes_{i=1}^{N} U(\theta) \ket{0}_i, \, \theta \in
	  \Theta \right\}.
	  \end{equation}
\end{corollary}
Note that the foregoing argument for ${\mathscr D}$ is meaningful only if we employ a fixed set of bases in all 
constituent spaces  $\mathscr{H}_i$,  since otherwise we can always perform unitary transformations in each of the spaces so that any state $\ket{\psi} \in {\mathscr T}$ has the form in (\ref{std}).
The characterization of ${\mathscr D}$ by (\ref{std}) has been obtained earlier in \cite{DB07} for $n=2$, which is also presented in \cite{HMMOV08} for generic $n$ in a form equivalent to the latter half of statement (ii) of Theorem 1.

Our arguments so far can be extended to the braid group $B_N$ whose generators $\sigma_i$, $1\le i\le N-1$, satisfy
\be
\begin{split}
\sigma_i\sigma_{i+1}\sigma_i=\sigma_{i+1}\sigma_i\sigma_{i+1}
\quad
&{\rm for}
\quad
i=1,2,\ldots,N-2,\\
\sigma_i\sigma_j=\sigma_j\sigma_i
\quad
&{\rm for}
\quad
|i-j|\ge2.\\
\end{split}
\ee
Indeed, if a state $\ket{\psi}$ follows some one dimensional representations of $B_N$, we have
\be
\pi_{\sigma_i}\ket{\psi}={\rm e}^{{\rm i}\chi}\ket{\psi},
\quad
\forall\,\sigma_i\in B_N,
\label{anyon}
\ee
with $\pi_{\sigma_i}$ appropriately defined for the braid group, 
for which statements analogous to those for fermions hold.  
This implies that states obeying intermediate symmetries ({\it i.e.}, ${\rm e}^{{\rm i}\chi}\neq \pm 1$), which correspond to
states realized by anyons in particle systems,  will always be globally entangled.

\section{Full Separability and Symmetry}
\setcounter{equation}{0}

In this section, we focus on symmetric states and discuss their separability further.   For this, we first introduce
the total symmetrization operator \cite{fn1},
    \begin{equation}
     \label{sym01}
      T:= {1\over{N !}} \sum_{\sigma\in S_N} \pi_{\sigma}.
    \end{equation}
Since $\pi_{\sigma} T = T$ for all $\sigma \in S_N$, we see that for
any $\ket{\psi} \in \mathscr{H}$
the totally symmetrized state $T\ket{\psi}$  is invariant under the action of the symmetric group, 
\begin{equation}
    \pi_{\sigma} T\ket{\psi} = T\ket{\psi},\qquad \forall\,\sigma\in S_N.
\end{equation}
Note that $T\ket{\psi} \in \mathscr{S}$, and conversely, if $\ket{\psi} \in \mathscr{S}$ then $\ket{\psi} = T\ket{\psi}$.  This implies
that the space $\mathscr{S}$ of symmetric states (\ref{symdef}) can also be written as
$\mathscr{S} = \{\, \ket{\psi} \; \big\vert \; \ket{\psi} = T\ket{\phi},\, \exists\,\ket{\phi} \in \mathscr{H} \}$.  

If we let $\mathscr{S}^\perp$ be the space of states which are orthogonal to all symmetric states, 
we can show
\begin{proposition}
A fully separable state cannot be contained in $\mathscr{S}^\perp$, {\it i.e.}, 
 \begin{equation}
\ket{\psi} \in \mathscr{T} \quad  \Rightarrow \quad  \ket{\psi} \not\in \mathscr{S}^\perp.
\end{equation}
\end{proposition}
\begin{proof}
To the contrary of the statement, assume that there exists a fully separable state $\ket{\psi}\in \mathscr{S}^\perp$.
Then $\braket{\phi|\psi}=0$ for all $\ket{\phi} \in \mathscr{S}$.
As before, we expand $\ket{\psi}$ as
	      \begin{equation}
	       \ket{\psi}=\bigotimes_{i=1}^N
		\left(\sum_{a_i=0}^{n-1} c^i_{a_i}\ket{a_i}_i\right),
		\qquad
c_{a_{i}}^i \in \mathbb{C}.
	      \end{equation}
Choose first  $\ket{\phi_0}=\ket{0}^{\otimes N} \in \mathscr{S}$ where $\ket{0}^{\otimes N}:= \otimes_{i=1}^N\ket{0}_i$ for which $\braket{\phi_0|\psi}=0$.  This is equivalent to
	      \begin{equation}
	       \prod_{i=1}^N c^i_0=0,
	      \end{equation}
which shows that at least one of the elements of \{$c^i_0\, |\, 1\le i\le N$\}  is zero.
For convenience, we renumber the indices $i = 1, \ldots, N$ of the constituents $\mathscr{H}_i$ so that 
\begin{equation}
\label{sym02}
	       c^i_0=0,
	       \quad
	       1\le i\le n_0
\end{equation}
holds for some $1 \leq n_0 \leq N$.
Next we choose
\begin{equation}
	       \ket{\phi_1}=T
	       \left(\ket{1}_1 \otimes \cdots \otimes
		\ket{1}_{n_0}\otimes \ket{0}_{n_0+1} \otimes \cdots
		\otimes \ket{0}_{N} \right) 
		\in \mathscr{S}.
\end{equation}
From $\braket{\phi_1|\psi}=0$ and (\ref{sym02}), we obtain
\begin{equation}
	       \prod_{i=1}^{n_0}c^i_1=0.
\end{equation}
Since $n_0 \geq 1$, at least one of the elements of \{$c^i_1\, |\, 1\le i\le n_0$\} is zero.
We renumber the indices again so that 
	     \begin{equation}
\label{sym03}
	       c^i_1=0,
	       \quad
	       1\le i\le n_1
\end{equation}
holds for some  $1\leq n_1 \leq n_0$.
Clearly, this process can be repeated until we arrive at
	      \begin{equation}
	      c^1_{a_1}=0,
	      \quad
	      0\le a_1\le n-1.
	      \end{equation}
This shows that $\ket{\psi} = 0$, invalidating the assumption made at the beginning.
\end{proof}

We remark that the contraposition of Proposition 2 is
 \begin{equation}
\ket{\psi} \in \mathscr{S}^\perp \quad  \Rightarrow \quad  \ket{\psi} \not\in \mathscr{T}.
\end{equation}
This shows that any state orthogonal to $\mathscr{S}$ cannot be fully separable,  which is an extension of the previous result  \cite{DB07} for general $n$.

In passing, we note that since $T$ is a projection operator $T^2 = T$, it can be used for the direct sum decomposition
of the Hilbert space $\mathscr{H} = \mathscr{S} \oplus \mathscr{S}^\perp$.  Namely, for any $\ket{\psi} \in \mathscr{H}$, we have
\begin{equation}
\label{decomp}
\ket{\psi} = \ket{\xi}  + \ket{\varphi} , \qquad \ket{\xi} \in \mathscr{S}, \quad \ket{\varphi} \in \mathscr{S}^\perp,
\end{equation}
with
\begin{equation}
\label{decomp}
\ket{\xi} = T \ket{\psi} , \qquad \ket{\varphi} = (1-T) \ket{\psi}.
\end{equation}
We also add that, if we have local unitary transformations $U(\theta^i)$ in $\mathscr{H}_i$ 
with common parameters $\theta^i = \theta \in \Theta$ for all $i$ and consider the combined unitary transformation 
$V(\theta) := U(\theta)\otimes \cdots \otimes U(\theta)$ in $\mathscr{H}$, the two operators $V$ and $T$ commute, 
\begin{equation}
\label{utcommute}
VT = TV.
\end{equation}

We then show
\begin{proposition}
If $\ket{\psi}$ is  a fully separable state whose projection $T\ket{\psi}$ on $\mathscr{S}$ is still fully separable, then $\ket{\psi}$ is symmetric, {\it i.e.}, 
 \begin{equation}
\ket{\psi} \in \mathscr{T}   \quad \hbox{\rm and} \quad  T \ket{\psi}\in\mathscr{T}
\quad  \Rightarrow \quad  \ket{\psi} \in \mathscr{S} .
\end{equation}
\end{proposition}
\begin{proof}
Let $\ket{\psi}$ be a fully separable state $\ket{\psi} \in \mathscr{T}$ for which 
$T \ket{\psi} \in \mathscr{T}$.   This  means $T \ket{\psi} \in \mathscr{D}$, and hence from Corollary 2 we have
$T \ket{\psi} = V(\theta)\ket{0}^{\otimes N}$ for some $\theta \in \Theta$.  
Since $T$ and $V$ commute as noted in (\ref{utcommute}), it follows that the state
\begin{equation}
\label{etast}
 \ket{\eta} :=  V^{-1}(\theta)\ket{\psi}
\end{equation}
generated from $\ket{\psi}$ by applying the inverse operator $V^{-1}$ of $V$
has the projection,
\be
T  \ket{\eta} = \ket{0}^{\otimes N}.
\label{tetast}
\ee
We then see that since $V^{-1}$ is a product of local unitary transformations, and since $\ket{\psi} \in \mathscr{T}$, the state
$\ket{\eta}$ in (\ref{etast}) is also fully separable $\ket{\eta} \in \mathscr{T}$.  
Now, from ${^{N\otimes}}\bra{0}T\ket{\eta}=1$ we may write $\ket{\eta}$ without loss of generality as
\begin{equation}
\label{etaexp}
 \ket{\eta}=\bigotimes_{i=1}^N
  \left(\ket{0}_i+\sum_{a=1}^{n-1}
   c^i_a\ket{a}_i
  \right).
\end{equation}
The projection of the state then reads
\begin{eqnarray}
 T \ket{\eta} &=& \ket{0}^{\otimes N}+ h_1 T(\ket{1}_1\ket{0}_2\ket{0}_3\cdots\ket{0}_{N-1}\ket{0}_N)
  +h_2 T(\ket{1}_1\ket{1}_2\ket{0}_3\cdots\ket{0}_{N-1}\ket{0}_N)+\cdots 
  	  \nonumber\\
&& {} \, + h_{N-1} T(\ket{1}_1\ket{1}_2\ket{1}_3\cdots\ket{1}_{N-1}\ket{0}_N)
+ h_{N} T(\ket{1}_1\ket{1}_2\ket{1}_3\cdots\ket{1}_{N-1}\ket{1}_N) +\cdots,
\label{etexp}
\end{eqnarray}
where the coefficients in the expansion (\ref{etexp}) can be determined from (\ref{etaexp}).  For instance, we find
\begin{equation}
 h_1= \sum_{k=1}^N c^k_1,\s
  h_2= \sum_{k,k'= 1, k\neq k'}^N c^k_1c^{k'}_1,  
\end{equation}
and
\begin{equation}
h_{N-1}= \sum_{k=1}^N c^1_1c^2_1\cdots \widehat{c^k_1} \cdots c^N_1,\s
 h_N=  \prod_{i=1}^Nc^i_1, 
\end{equation}
where the hat in $\widehat{c^k_1}$ indicates that the factor ${c^k_1}$ is missing in the product.
To be consistent with (\ref{tetast}), all of these coefficients must vanish.  From $h_N = 0$, we see that
at least one of the elements of $\{c^i_1\, |\, 1\le i\le N\}$ is zero, and we may choose $c^1_1 = 0$ for definiteness.  Then from $h_{N -1}= 0$ we find that at least one of the elements of $\{c^i_1\, |\, 2\le i\le N\}$ is zero, and again we choose $c^2_1 = 0$.  This procedure can be repeated until we reach $h_1 = 0$ which yields $c^N_1 = 0$ and, consequently, we obtain
 \begin{equation}
	      c^i_1=0,
	      \quad
	      1\le i\le N.
	      \end{equation}
The conclusion will not be changed even if we choose different factors for the vanishing elements in each of the steps in the procedure.   Obviously, similar arguments can be applied for the terms  
$ T(\ket{i}_1\ket{0}_2\cdots\ket{0}_N)$ up to $T(\ket{i}_1\ket{i}_2\cdots\ket{i}_N)$ for $i \ge 2$ as well.  Combining all the results obtained at the end of these procedures, we find
\begin{equation}
c^i_{a}=0,
\qquad 
1\le i\le N,
\quad
0\le a \le n-1.
\end{equation}
This shows that $\ket{\eta} = \ket{0}^{\otimes N}$, which in turn indicates 
that $\ket{\psi} = U \ket{\eta} \in \mathscr{S}$,  proving Proposition 3.
\end{proof}

We remark that Proposition 3 can equally be expressed as
\begin{equation}
\ket{\psi} \in \mathscr{T}   \quad \hbox{and} \quad \ket{\psi} \not \in \mathscr{S} \quad  
\Rightarrow \quad  T \ket{\psi} \not\in \mathscr{T},
\end{equation}
which implies that it is impossible to retain the full separability of a state 
under total symmetrization unless the original state is symmetric.

\section{Conclusion and Discussions}
In this paper, we studied some of the basic properties of entanglement of multipartite systems based on exchange symmetry for general
constituent number $N$ and also for general level $n$.   
The states primarily considered are 
those which are symmetric or anti-symmetric under permutations of the constituent states comprising the system.    
We argued that different partitions of a state are mutually compatible (Lemma 1), and that symmetric property under a permutation requires a certain rigid structure for the state imposed by the action of the permutation on the constituent systems (Lemma 2).   Based on these observations we arrive at Theorem 1, which states that if a state is symmetric under all transpositions, it is either globally entangled or otherwise it is fully separable, sharing essentially the same constituent states.   Further, if the state is anti-symmetric, it is necessarily globally entangled.    
We also presented a number of  propositions revealing the close connection between symmetry and separability, some of which are a generalization of the recent results of Ref.\cite{DB07} for multipartite $n$-level systems.  

We remark that the statement of Theorem 1 can be extended to states which are anyonic under transpositions, that is, like anti-symmetric states,  anyonic states with eigenvalue 
$\lambda = {\rm e}^{{\rm i}\chi} \ne \pm 1$ must always be globally entangled.    If our arguments can be generalized to 
systems with an infinite dimensional Hilbert space $\mathscr{H}$, our results may also be useful for analyzing physical properties of entangled states in, {\it e.g.}, infinite spin systems or particle systems moving on a plane where anyonic states are permitted.  In this respect, it will be of interest to investigate the possible role of entanglement in the manifestation of phase transitions or topological quantizations observed in these systems.   In fact, for systems of $N$-qubits the behavior of bipartite entanglement in symmetric states under phase transitions has been studied in \cite{PS05, Vidal06, VDB07}.  

Our results show that exchange symmetry 
implies that states are globally entangled unless they are fully separable, but it does not tell how globally entangled they are, if not separable.   This is an important question because it is known that globally entangled states can further be classified
based on stochastic local operations assisted with classical communications (SLOCC) \cite{DVC00, VDMV02}.  In the case of 3-qubit systems, for instance, the GHZ state and the W state are shown to be representatives of two classes of globally entangled states which have distinct physical properties.    Technically, the finer classification of globally entangled states is carried out in terms of ranks and ranges of the reduced density matrices, which are also symmetric under exchange when the original states are. 
Thus, exchange symmetry  {\it per se} is not useful for the classification, but it may be possible to use it to narrow down the states under consideration for the classification, finding the representatives of distinct classes as the GHZ and the W states which are both symmetric.

The intimate link between symmetry and entanglement found in this paper suggests that there can be a novel account of  known physical properties from exchange symmetry of entangled states.    This line of study will be facilitated if we can 
extend our arguments to mixed states with exchange symmetry, as has been done for the the case of bipartite systems in \cite{ESBL02,NV07} based on approaches specific to indistinguishable particles \cite{SCKLL01}. 
As a first step, in the Appendix we present an extension of Lemma 1 to mixed states which are separable in a
restricted class.
As in the case of pure states, it is also important to study the possibility of classifying globally
entangled mixed states using exchange symmetry, possibly along with the witness operators \cite{ACIN01}.
Finally, we mention that our arguments in this paper are given solely on the basis of states and do not  rely on any particular entanglement measures.   Thus, our results may also serve as a testing ground to find better entanglement measures in order  to characterize the interplay  of multipartite constituent states more comprehensively.

\begin{acknowledgments}
This work has been supported in part by
the Grant-in-Aid for Scientific Research (C), No.~20540391, of
the Japanese Ministry of Education, Science, Sports and Culture.
\end{acknowledgments}

 \renewcommand{\theequation}{A.\arabic{equation}}
 \setcounter{equation}{0}  
 \section*{APPENDIX}  
  
In this Appendix we 
show that Lemma 1 can be extended to a certain type of separable mixed states.  
To be specific, let us call a mixed state 
$\rho$ {\it directly separable w.r.t.~$\mathscr{A}$}, if it admits the decomposition
$\rho=\bigotimes_{k=1}^s\rho_{\alpha_k}$ according to the partition
$\mathscr{A}= \{\mathscr{A}_k\}_{k=1}^s$.   
Clearly, these mixed states are a direct extension of the separable pure states (\ref{statedp}) and form a restricted class of the general separable mixed states \cite{RFW89}.  We then have
\begin{lemma}
 If a density matrix $\rho$ is directly separable w.r.t.~two partitions $\mathscr{A}=\{\mathscr{A}_k\}_{k=1}^s$ and $\mathscr{B}=\{\mathscr{B}_l\}_{l=1}^t$, {\it i.e.}, 
\begin{equation}
 \rho =\bigotimes_{k=1}^s \rho_{\alpha_k} = \bigotimes_{l=1}^t \rho_{\beta_l},
\end{equation}
then, the density matrix is also directly separable w.r.t.~the partition $\mathscr{C}=\mathscr{A}\cap\mathscr{B}$, that is, 
it can be written as
\begin{equation}
 \rho = \bigotimes_{k=1}^s\bigotimes_{l=1}^t \rho_{\alpha_k\cap\beta_l}.
\end{equation}
\end{lemma}
\begin{proof}
We just use the assumptions to obtain
\begin{eqnarray}
 \rho &=& \bigotimes_{k=1}^s\rho_{\alpha_k} =  \bigotimes_{k=1}^s\Tr_{\bar{\alpha}_k}\rho
 =
  \bigotimes_{k=1}^s\bigotimes_{l=1}^t\Tr_{\bar{\alpha}_k}\rho_{\beta_l}\nonumber\\
 &=&
 \bigotimes_{k=1}^s\bigotimes_{l=1}^t\Tr_{\bar{\alpha}_k\cup\bar{\beta}_l}\rho
 =
 \bigotimes_{k=1}^s\bigotimes_{l=1}^t\Tr_{\overline{\alpha_k\cap\beta_l}}\rho
 =
  \bigotimes_{k=1}^s\bigotimes_{l=1}^t\rho_{\alpha_k\cap\beta_l},
\end{eqnarray}
which proves the statement.
\end{proof}

Using this lemma, and considering  mixed states with symmetries under exchange, one may further arrive at a theorem analogous to Theorem 1.    It seems, however, that it requires some additional ingredients to extend our arguments to the most general class of separable mixed states which admit more involved probabilistic structures than those realized by
the pure states and the class of mixed states discussed above.



\begin{thebibliography}{99}
\bibitem{fn1}
Note that the action of the total symmetrization operator $T$ is in general basis-dependent, because the permutations  (\ref{permac}) are performed among the corresponding basis states 
chosen in each of  $\mathscr{H}_i$.
\end{thebibliography}
\end{document}